\documentclass{article}
\begin{document}


\title{Logical equivalence between generalized urn models and finite automata}
\author{Karl Svozil\\
 {\small Institut f\"ur Theoretische Physik, University of Technology Vienna }     \\
  {\small Wiedner Hauptstra\ss e 8-10/136,}
  {\small A-1040 Vienna, Austria   }            \\
  {\small e-mail: svozil@tuwien.ac.at}}

\maketitle
\begin{abstract}
To every generalized urn model there exists a finite (Mealy) automaton
with identical propositional calculus.
The converse is true as well.
\end{abstract}


\section{Introduction of concepts}

In what follows we shall explicitly and constructively demonstrate
the equivalence of the empirical logics
(i.e., the propositional calculi) associated with the
generalized urn models (GUM) suggested by
Ron Wright \cite{wright:pent,wright}, and automaton partition logics
(APL)
\cite{svozil-93,schaller-96,dvur-pul-svo,cal-sv-yu,svozil-ql}.
(The result has been  mentioned already in \cite[p.145]{svozil-ql}, but no proof has been given).
The logical equivalence of automaton models (AM)
with generalized urn models suggests that these logics are more general and ``robust''
with respect to changes of the particular model than could have been expected from
the particular instances of their first appearance.

Wright's original generalized urn model has been designed to conform to the Foulis-Randal setup,
such that color independence is implemented in the following form
[cf. the ``constancy axiom'' (ii) on \cite[p. 889]{wright}]:
if  a particular symbol can occur in two operations, it means the same in both.
Stated differently, if the same symbol appears in two different colors, then any ball
on which it appears in one color must have it in the other color as well.
We shall consider a modification of this generalized urn model: in
our case, the same symbol in different colors need not match on the same ball; i.e.,
the same symbols in different colors may also appear on different balls.

\subsection{Generalized urn models}
A generalized urn model
${\cal U}=\langle U,C,L,\Lambda \rangle $ is
characterized as follows.
Consider an ensemble of balls with black background color.
Printed on these balls are some color symbols from a symbolic alphabet $L$.
The colors are elements of a set of colors $C$.
A particular ball type is associated with a unique combination of mono-spectrally
(no mixture of wavelength) colored symbols
printed on the black ball background.
Let $U$ be the set of ball types.
We shall assume that every ball contains
just one single symbol per color.
(Not all types of balls; i.e., not all color/symbol combinations, may be present in
the ensemble, though.)

Let
$\vert U\vert $ be the number of different types of balls,
$\vert C\vert $ be the number of different mono-spectral colors,
$\vert L\vert $ be the number of different output symbols.

Consider the deterministic ``output'' or ``lookup''
function $\Lambda (u,c)=v$,
$u\in U$,
$c\in C$,
$v\in L$,
which returns one symbol per ball type and color.
One interpretation of this lookup function $\Lambda$ is as follows.
Consider a set of $\vert C\vert $ eyeglasses build from filters for the
$\vert C\vert $ different colors.
Let us assume that these mono-spectral filters are
``perfect'' in that they totally absorb light of all other colors
but a particular single one.
In that way, every color can be associated with a particular eyeglass and vice versa.

When a spectator looks at a particular ball through such an eyeglass,
the only operationally recognizable symbol will be the one in the particular
color which is transmitted through the eyeglass.
All other colors are absorbed, and the symbols printed in them will appear black
and therefore cannot be differentiated from the black background.
Hence the ball appears to carry a different ``message'' or symbol,
depending on the color at which it is viewed.

An empirical logic can be constructed as follows.
Consider the set of all ball types.
With respect to a particular colored eyeglass, this set disjointly ``decays''
or gets partitioned into those ball types which can be separated by the particular color of
the eyeglass.
Every such partition of ball types can then be identified with a Boolean algebra whose atoms are the elements of the partition.
A pasting of all of these Boolean algebras yields the empirical logic associated with
the particular urn model.

\subsection{Automaton models}
A (Mealy type) automaton
${\cal A}=\langle S,I,O,\delta ,\lambda \rangle$ is characterized
by the set of states $S$,
by the set of input symbols $I$,
and by the set of output symbols $O$.
$\delta (s,i)=s'$ and
$\lambda (s,i)=o$,
$s,s'\in S$,
$i\in I$
and $o\in O$
represent the transition and the output functions, respectively.
The restriction to Mealy automata is for convenience only.

A typical automaton experiment aims at
an operational determination of an unknown initial state
by the input of some symbolic sequence and the observation of the resulting output symbols.
Every such input/output experiment results
in a state partition in the following way.
Consider a particular automaton.
Every experiment on such an automaton which tries to
solve the initial state problem is characterized
by a set of input/output symbols
as a result of the possible input/output sequences for this experiment.
Every such distinct set of input/output symbols is associated with a set of initial automaton
states which would reproduce that sequence.
This state set may contain one or more states, depending on the ability of the experiment
to separate different initial automaton states.
A partitioning of the automaton states is obtained if one considers
a single input sequence and the variety of all
possible output sequences (given a particular automaton).
Stated differently:
given a set of inputs, the set of automaton states decays into disjoint
subsets associated with the possible output sequences.
(All elements of a subset yield the same output on the same input.)

This partition can then be identified with a Boolean algebra,
with the elements of the partition interpreted as atoms.
By pasting the Boolean algebras of the ``finest'' partitions together one obtains
an empirical partition logic associated with the particular automaton.
(The converse construction is also possible, but not unique; see below.)

For the sake of simplicity,
we shall assume that every experiment just deals with a single input/output
combination.
That is, the finest partitions are reached already after the first symbol.
This does not impose any restriction on the partition logic, since
given any particular automaton, it is always possible
to construct another automaton with exactly the same partition logic as the first one
with the above property.

More explicitly, given any partition logic, it is always possible to
construct a corresponding automaton with the following specification:
associate with every element of the set of partitions a single input symbol.
Then take the partition with the highest number of elements and associate a single output
symbol with any element of this partition.
(There are then sufficient output
symbols available for the other partitions as well.)
Different partitions require different input symbols;
one input symbol per partition.
The output function  can then be defined by associating a single output symbol per element
of the partition (associated with a particular input symbol).
Finally, choose a transition function which completely looses the state information
after only one transition; i.e., a transition function which maps all automaton state into
a single one.

\section{Proof of logical equivalence}

From the definitions and constructions mentioned in the previous sections
it is intuitively clear
that, with respect to the empirical logics,
generalized urn models and finite automata models
are equivalent.
Every logic associated with a generalized urn model
can be interpreted as an automaton partition logic
associated with some (Mealy) automaton (actually an infinity thereof).
Conversely, any logic associated with some (Mealy) automaton
can be interpreted as a logic associated with some generalized urn model
(an infinity thereof).
We shall proof these claims by explicit construction.
Essentially, the lookup function
$\Lambda$ and the output function $\lambda$ will be identified.
Again, the restriction to Mealy automata is for convenience only.
The considerations are robust with respect to variations of finite input/output automata.

\subsection{Direct construction of AM from GUM}

In order to define an APL associated with a Mealy automaton
${\cal A}=\langle S,I,O,\delta ,\lambda \rangle$
from a GUM ${\cal U}=\langle U,C,L,\Lambda \rangle $,
let
$u\in U$,
$c\in C$,
$v\in L$,
and
$s,s'\in S$,
$i\in I$,
$o\in O$, and assume
$\vert U\vert =\vert S\vert$,
$\vert C\vert =\vert I\vert$,
$\vert L\vert =\vert O\vert$.
The following identifications can be made
with the help of the bijections $t_S,t_I$ and $t_O$:
\begin{equation}
\begin{array}{llllll}
t_S(u)=s, \;
 t_I(c)=i, \;
 t_O(v)=o, \\
\delta (s,i)= s_i  \quad {\rm for \; fixed\;}s_i\in S {\rm \;and \;
arbitrary\;}s\in S,\; i\in I,\\
\lambda  (s,i) = t_O\left(\Lambda (t_S^{-1}(s),t_I^{-1}(i))\right).
\end{array}
\label{oto-ua}
\end{equation}
More generally, one could use equivalence classes instead of a bijection.
Since the input-output behavior is equivalent and the
automaton transition function is trivially $\vert L\vert $-to-one,
both entities yield the same propositional calculus.

\subsection{Direct construction of GUM from AM}

Conversely,
consider an arbitrary Mealy automaton ${\cal A}=\langle S,I,O,\delta ,\lambda \rangle$
and its associated propositional calculus APL.

Just as before, associate with every single automaton state $s\in S$
a ball type $u$,
associate with every input symbol $i\in I$
a unique color $c$,
and
associate with every output symbol $o\in O$
a unique symbol $v$; i.e., again
$\vert U\vert =\vert S\vert$,
$\vert C\vert =\vert I\vert$,
$\vert L\vert =\vert O\vert$.
The following identifications can be made
with the help of the bijections $\tau_U,\tau_C$ and $\tau_L$:
\begin{eqnarray}
\begin{array}{llllll}
\tau_U(s)=u, \; \tau_C(i)=c, \;  \tau_L(o)=v, \;
\Lambda  (u,c) = \tau_L (\lambda (\tau_U^{-1}(u),\tau_C^{-1}(c))).
\end{array}
\label{oto-au}
\end{eqnarray}
A direct comparison of
(\ref{oto-ua})
and
(\ref{oto-au})
yields
\begin{eqnarray}
\begin{array}{llllll}
\tau_U^{-1}=t_S, \; \tau_C^{-1}=t_I, \;  \tau_L^{-1}=t_O.
\end{array}
\label{oto-oto}
\end{eqnarray}

\subsection{Schemes using dispersion-free states}

Another equivalence scheme  uses the fact that both
automaton partition logics and the logic of generalized urn models
have a separating (indeed, full) set of
dispersion-free states.
(In what follows, the terms
``dispersion-free state''
``two-valued state''
``valuation''
``dispersion-free probability measure''
are  synonyms for measures which take on only the values zero and one.
We thereby explicitly exclude dispersion-free measures which take on other values, such as
$1/2$ and $0$, as introduced by Wright \cite{wright:pent}.)
Stated differently, given a finite atomic logic with a separating set of states, then
the enumeration of the complete set of dispersion-free states enables
the explicit construction of  generalized urn models and automaton logics
whose logic corresponds to the original one.

This can be achieved by ``inverting'' the set of two-valued states as follows.
(The method is probably best understood by considering the examples below.)
Let us start with an atomic logic with a separating set of states.
\begin{itemize}
\item[(i)]
In the first step, every atom of this lattice is labeled by some natural number,
starting from ``$1$'' to ``$n$'', where $n$  stands for the number of lattice atoms.
The set of atoms is denoted by $A=\{1,2,\ldots , n\}$.

\item[(ii)]
Then, all two-valued states of this lattice are labeled consecutively
by natural numbers, starting from ``$m_1$'' to ``$m_r$'', where $r$  stands for the number of
two-valued states.
The set of states is denoted by $M=\{m_1,m_2,\ldots , m_r\}$.

\item[(iii)]
Now  partitions are defined as follows.
For every atom, a set is created whose members are the numbers or ``labels'' of
the two-valued states which are ``true'' or take on the value ``1'' on this atom.
More precisely,
the elements $p_i(a)$ of the partition ${\cal P}_j$ corresponding to
some atom $a\in A$ are defined by
$$p_i (a) =
\left\{
k \mid m_k(a)=1, \; k\in M
\right\}
.
$$
The partitions are obtained by taking the unions of all $p_i$ which belong to the same
subalgebra ${\cal P}_j$.
That the corresponding sets are indeed partitions follows from the properties of
two-valued states: two-valued states (are ``true'' or) take on the value ``$1$'' on just one atom
per subalgebra and (``false'' or) take on the value ``$0$'' on all other atoms of this subalgebra.

\item[(iv)]
Let there be $t$ partitions labeled by ``1'' through ``$t$''.
The partition logic is obtained by a pasting of all partitions
${\cal P}_j$, $1\le j \le t$.

\item[(v)] In the following step, a corresponding GUM or automaton model is
obtained from the partition logic just constructed.

\begin{itemize}
\item[(a)] A GUM is obtained by the following identifications
(see also \cite[p. 271]{wright:pent}).
\begin{itemize}
\item[$\bullet$]
Take as many ball types as there are two-valued states; i.e.,
$r$ types of balls.
\item[$\bullet$]
Take as many colors as there are subalgebras or partitions; i.e., $t$ colors.
\item[$\bullet$]
Take as many symbols as there are elements in the partition(s) with the maximal number of elements;
i.e., $\max_{1\le j\le t}\vert {\cal P}_j\vert \le n$.
To make the construction easier, we may just take as many symbols as there are atoms; i.e., $n$ symbols.
(In most cases, much less symbols will suffice).
Label the symbols by $v_l$.
Finally, take $r$ ``generic'' balls with black background.
Now associate with every measure a different ball type.
(There are $r$ two-valued states, so there will be $r$ ball types.)
\item[$\bullet$]
The $i$th ball type is painted by colored
symbols  as follows:
Find the atoms  for which the $i$th two-valued state $m_i$ is $1$.
Then paint the symbol corresponding to every such lattice atom on the ball, thereby choosing
the color associated with the subalgebra or partition
the atom belongs to.
If the atom belongs to more than one subalgebra,
then paint the same symbol in as many colors as there are partitions or subalgebras
the atom belongs to (one symbol per subalgebra).
\end{itemize}
This completes the construction.

\item[(b)] A Mealy automaton is obtained by the following identifications
(see also \cite[pp. 154--155]{svozil-93}).
\begin{itemize}
\item[$\bullet$]
Take as many automaton states as there are two-valued states; i.e.,
$r$  automaton states.
\item[$\bullet$]
Take as many input symbols as there are subalgebras or partitions; i.e., $t$ symbols.
\item[$\bullet$]
Take as many output symbols as there are elements in the partition(s) with the maximal number of elements
(plus one additional auxiliary output symbol ``$\ast$'', see below);
i.e., $\max_{1\le j\le t}\vert {\cal P}_j\vert \le n+1$.
\item[$\bullet$]
The output function is chosen to match the elements of the state partition corresponding
to some input symbol.
Alternatively, let the lattice atom $a_q\in A$
must be an atom of the subalgebra corresponding to the input $i_l$.
Then one may choose an output function such as
$$
\lambda (m_k,i_l)= \left\{
\begin{array}{l}
a_q \quad{\rm if }\;m_k (a_q)= 1\\
\ast \; \quad {\rm if }\;m_k (a_q)= 0\\
\end{array}
\right.
$$
with
$1\le k \le r$
and
$1\le l \le t$.
Here, the additional output symbol ``$\ast$'' is needed.

\item[$\bullet$]
The transition function is $r$--to--1 (e.g., by $\delta (s,i)=s_1$, $s,s_1\in S$,
$i\in I$), i.e., after one input the information about the
initial state is completely lost.
\end{itemize}
This completes the construction.
\end{itemize}
\end{itemize}

\subsection{Example 1: The generalized urn logic $L_{12}$}

In what follows we shall illustrate the above constructions with a couple of examples.
First, consider the generalized urn model
$$\langle \{u_1,\ldots ,u_5\},\{{\rm red,green}\},\{1,\ldots ,5\},\Lambda \rangle$$
with $\Lambda$ listed in Table \ref{t-wright}(a).
\begin{table}
\begin{center}
\begin{tabular}{ll}
\begin{tabular}{|c|cc|}
\hline\hline
ball type & red & green \\ \hline
1  & $1$ & $3$ \\
2  & $1$ & $4$ \\
3  & $2$ & $3$ \\
4  & $2$ & $4$ \\
5  & $5$ & $5$ \\
\hline\hline
\end{tabular}
\qquad
\qquad
&
\begin{tabular}{|c|ccccc|ccccc|}
 \hline\hline
 &&&$\delta$ && & &&$\lambda$&&\\
\raisebox{2.5ex}[0cm][0cm]{state} &1&2&3&4&5 & 1&2&3&4&5\\
 \hline
0&1&1&1&1&1 & 1&1&2&2&5\\
1&1&1&1&1&1 & 3&4&3&4&5\\
 \hline\hline
\end{tabular}
\\
(a)&(b)\\
\end{tabular}
\end{center}
\caption{\label{t-wright} (a) Ball types in Wright's generalized urn model
\protect\cite{wright} (cf. also \protect\cite[p.143ff]{svozil-ql}).
(b) Transition and output table of an associated automaton model.}
\end{table}

The associated Mealy automaton can be directly constructed as follows.
Take $t_S=t_O= {\rm id}$, where $ {\rm id}$ represents the identity function,
and take
$t_I({\rm red})=0$
and
$t_I({\rm green})=1$,
respectively.
Furthermore, fix a (five$\times$two)-to-one transition function by $\delta(.,.)=1$.
The transition and output tables are listed in Table  \ref{t-wright}(b).
Both empirical structures yield the same propositional logic $L_{12}$.

\subsection{Example 2: The automaton partition logic $L_{12}$}
Let us start with
an automaton whose transition and output tables are listed in Table  \ref{t-wright}(b)
and indirectly construct a logically equivalent GUM by using dispersion-free states.
The first thing to do is to figure out all dispersion-free states of $L_{12}$.
There are five of them, which we might write in vector form; i.e., in lexicographic order:
\begin{equation}
\begin{array}{c}
m_1= (0,0,0,0,1),\;
m_2= (0,1,0,1,0),\;
m_3= (0,1,1,0,0), \\
m_4= (1,0,0,1,0), \;
m_5= (1,0,1,0,0).
\end{array}  \nonumber
\end{equation}

Now define the following GUM as follows.
There are two subalgebras with the atoms $1,2,5$ and $3,4,5$, respectively.
Since there are five two-valued measures corresponding to five ball types.
They are colored according to the coloring rules defined above.
and $\Lambda$ as listed in Table
\ref{t-wright1}.
\begin{table}
\begin{center}
\begin{tabular}{|c|ccccc|ccccc|}
\hline\hline
&\multicolumn{10}{c|}{colors}\\
& \multicolumn{5}{c}{$c_1$}  & \multicolumn{5}{c|}{$c_2$}   \\
\raisebox{2.5ex}[0cm][0cm]{ball type} & \multicolumn{5}{c}{``red''}& \multicolumn{5}{c|}{``green''}   \\ \hline
1  & $\ast $& $\ast $ &$\ast $&$\ast$& 5      & $\ast $&$\ast$& $\ast $& $\ast$ & 5 \\
2  & $\ast $& 2       &$\ast $&$\ast$& $\ast$ & $\ast $&$\ast$& $\ast $& 4      & $\ast$ \\
3  & $\ast $& 2       &$\ast $&$\ast$& $\ast$ & $\ast $&$\ast$& 3      & $\ast $& $\ast$ \\
4  & 1      & $ \ast $&$\ast $&$\ast$& $\ast$ & $\ast $&$\ast$& $\ast$ & 4      & $\ast$ \\
5  & 1      & $ \ast $&$\ast $&$\ast$& $\ast$ & $\ast $&$\ast$& 3      & $\ast $& $\ast$ \\
\hline\hline
\end{tabular}
\end{center}
\caption{\label{t-wright1} Representation of the sign coloring scheme $\Lambda$.
``$\ast$'' means no sign at all (black) for the corresponding atom.}
\end{table}

\subsection{Example 3: GUM of the Kochen-Specker ``bug'' logic}
\begin{figure}
\begin{center}
\unitlength 0.85mm
\linethickness{0.4pt}
\begin{picture}(108.00,55.00)
\put(15.00,17.09){\circle{2.00}}
\put(25.00,7.33){\circle{2.00}}
\put(55.00,27.33){\circle{2.00}}
\put(85.00,7.33){\circle{2.00}}
\put(95.00,17.33){\circle{2.00}}
\put(25.00,7.33){\line(1,0){60.00}}
\put(25.00,47.33){\line(1,0){60.00}}
\put(55.00,7.33){\line(0,1){40.00}}
\put(25.00,7.33){\line(-1,1){20.00}}
\put(5.00,27.33){\line(1,1){20.00}}
\put(85.00,7.33){\line(1,1){20.00}}
\put(105.00,27.33){\line(-1,1){20.00}}
\put(24.67,55.00){\makebox(0,0)[rc]{$a_3=\{10,11,12,13,14\}$}}
\put(55.33,55.00){\makebox(0,0)[cc]{$a_4=\{2,6,7,8\}$}}
\put(85.33,55.00){\makebox(0,0)[lc]{$a_5=\{1,3,4,5,9\}$}}
\put(9.00,40.00){\makebox(0,0)[rc]{$a_2=\{4,5,6,7,8,9\}$}}
\put(99.33,40.00){\makebox(0,0)[lc]{$a_6=\{2,6,8,11,12,14\}$}}
\put(0.00,26.33){\makebox(0,0)[rc]{$a_1=\{1,2,3\}$}}
\put(108.00,26.33){\makebox(0,0)[lc]{$a_7=\{7,10,13\}$}}
\put(60.33,31.33){\makebox(0,0)[lc]{$a_{13}=$}}
\put(60.33,26.33){\makebox(0,0)[lc]{$\{1,4,5,10,11,12\}$}}
\put(9.00,13.33){\makebox(0,0)[rc]{$a_{12}=\{4,6,9,12,13,14\}$}}
\put(99.67,13.33){\makebox(0,0)[lc]{$a_8=\{3,5,8,9,11,14\}$}}
\put(24.67,-0.05){\makebox(0,0)[rc]{$a_{11}=\{5,7,8,10,11\}$}}
\put(55.33,-0.05){\makebox(0,0)[cc]{$a_{10}=\{3,9,13,14\}$}}
\put(85.33,-0.05){\makebox(0,0)[lc]{$a_9=\{1,2,4,6,12\}$}}
\put(5.00,27.33){\circle{2.00}}
\put(15.00,37.33){\circle{2.00}}
\put(25.00,47.33){\circle{2.00}}
\put(55.00,47.33){\circle{2.00}}
\put(85.00,47.33){\circle{2.00}}
\put(55.00,7.33){\circle{2.00}}
\put(104.76,27.33){\circle{2.00}}
\put(95.00,37.33){\circle{2.00}}
\end{picture}
\end{center}
\caption{\label{2001-cesena-f2} Greechie diagram of automaton partition logic
with a nonfull set of dispersion-free measures.}
\end{figure}
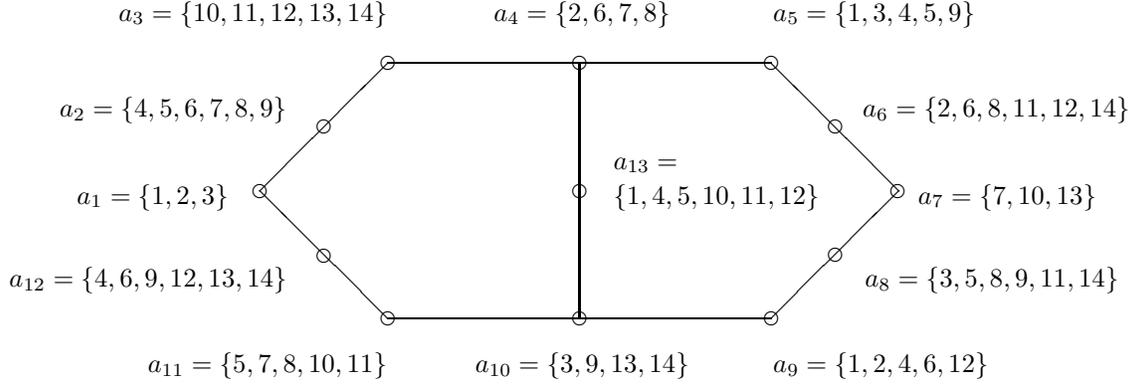

Another, less simple example, is a logic which is already mentioned by Kochen
and Specker \cite{kochen1} (this is a subgraph of their $\Gamma_1$)
whose automaton partition logic is depicted in Fig. \ref{2001-cesena-f2}.
(It is called ``bug'' by Professor Specker  \cite{Specker-priv} because of the similar shape with a bug.)
There are 14 dispersion-free states which are listed in Table \ref{2001-cesena-t2}(a).
The associated GUM is listed in Table \ref{2001-cesena-t2}(b).
\begin{table}
\normalsize
\begin{center}
{
\setlength{\tabcolsep}{3pt}
\begin{tabular}{|c|ccccccccccccc||ccccccc|}
\hline\hline
&\multicolumn{13}{c||}{lattice atoms}&\multicolumn{7}{|c|}{colors}\\
\raisebox{1.5ex}[0cm][0cm]{$m_r$ and}&$a_1$&$a_2$&$a_3$&$a_4$&$a_5$&$a_6$&$a_7$&$a_8$&$a_9$&$a_{10}$&$a_{11}$&$a_{12}$&$a_{13}$&$c_1$&$c_2$&$c_3$&$c_4$&$c_5$&$c_6$&$c_7$\\
\raisebox{1.5ex}[0cm][0cm]{ball type}&&&&&&&&&&&&&&&&&&&&\\
\hline
1  &1&0&0&0&1&0&0&0&1&0&0&0&1&  1&5&5&9&9 &1 &13        \\
2  &1&0&0&1&0&1&0&0&1&0&0&0&0&  1&4&6&9&9 &1 &4         \\
3  &1&0&0&0&1&0&0&1&0&1&0&0&0&  1&5&5&8&10&3 &10        \\
4  &0&1&0&0&1&0&0&0&1&0&0&1&1&  2&5&5&9&9 &12&13        \\
5  &0&1&0&0&1&0&0&1&0&0&1&0&1&  2&5&5&8&11&11&13        \\
6  &0&1&0&1&0&1&0&0&1&0&0&1&0&  2&4&6&9&9 &12&4         \\
7  &0&1&0&1&0&0&1&0&0&0&1&0&0&  2&4&7&7&11&11&4         \\
8  &0&1&0&1&0&1&0&1&0&0&1&0&0&  2&4&6&8&11&11&4         \\
9  &0&1&0&0&1&0&0&1&0&1&0&1&0&  2&5&5&8&10&12&10        \\
10 &0&0&1&0&0&0&1&0&0&0&1&0&1&  3&3&7&7&11&11&13        \\
11 &0&0&1&0&0&1&0&1&0&0&1&0&1&  3&3&6&8&11&11&13        \\
12 &0&0&1&0&0&1&0&0&1&0&0&1&1&  3&3&6&9&9 &12&13        \\
13 &0&0&1&0&0&0&1&0&0&1&0&1&0&  3&3&7&7&10&13&10        \\
14 &0&0&1&0&0&1&0&1&0&1&0&1&0&  3&3&6&8&10&12&10        \\
\hline\hline
\end{tabular}
}
\end{center}
\normalsize
\caption{Dispersion-free states  of the Kochen-Specker ``bug'' logic with 14 dispersion-free states.
and the associated GUM (all blank entries ``$\ast$''have been omitted).
\label{2001-cesena-t2}}
\end{table}

\section{Discussion}
We have explicitly demonstrated the logical equivalence of
generalized urn models and and the logic of finite
automata, both by a direct construction and by an indirect construction utilizing the
set of two-valued states.
This logical equivalence stresses the importance of these empirical structures.

GUMs and automata are capable to
serve as models for particular types of lattices with a sufficient number
of two-valued states (e.g.,  with a separating set of states).
Yet it is this very property which makes impossible the
realization of other, more exotic states, which
have no classical and not even a quantum mechanical counterpart.
Take, as an example,
the Wright state
\cite{wright:pent,svozil-ql}
on
the pentagon (or any $n$-agon, with odd $n>3$, $n=2k+1$, $k=2,3,\ldots$) Greechie diagram
with value $1/2$ on the five vertices and $0$ on each middle atom (three atoms per subalgebra).
The 11 two-valued measures suffice to generate GUMs and finite automata with that logical structure,
but none such model realizes the Wright state.

\section*{Acknowledgments}
This manuscript has been prepared on the request of a friend and colleague Norbert Brunner
who was considerably involved in the foundational stages of automaton
partition logic. I also kindly thank Ron Wright for his comments and suggestions,
in particular for pointing out the difference of his generalized urn model and the one considered here
with respect to the constancy axiom.


\begin{thebibliography}{10}

\bibitem{wright:pent}
Ron Wright.
\newblock The state of the pentagon. {A} nonclassical example.
\newblock In A.~R. Marlow, editor, {\em Mathematical Foundations of Quantum
  Theory}, pages 255--274. Academic Press, New York, 1978.

\bibitem{wright}
Ron Wright.
\newblock Generalized urn models.
\newblock {\em Foundations of Physics}, 20:881--903, 1990.

\bibitem{svozil-93}
Karl Svozil.
\newblock {\em Randomness \& Undecidability in Physics}.
\newblock World Scientific, Singapore, 1993.

\bibitem{schaller-96}
Martin Schaller and Karl Svozil.
\newblock Automaton logic.
\newblock {\em International Journal of Theoretical Physics}, 35(5):911--940,
  May 1996.

\bibitem{dvur-pul-svo}
Anatolij Dvure{\v{c}}enskij, Sylvia Pulmannov{\'{a}}, and Karl Svozil.
\newblock Partition logics, orthoalgebras and automata.
\newblock {\em Helvetica Physica Acta}, 68:407--428, 1995.

\bibitem{cal-sv-yu}
Cristian Calude, Elena Calude, Karl Svozil, and Sheng Yu.
\newblock Physical versus computational complementarity {I}.
\newblock {\em International Journal of Theoretical Physics}, 36(7):1495--1523,
  1997.

\bibitem{svozil-ql}
Karl Svozil.
\newblock {\em Quantum Logic}.
\newblock Springer, Singapore, 1998.

\bibitem{kochen1}
Simon Kochen and Ernst~P. Specker.
\newblock The problem of hidden variables in quantum mechanics.
\newblock {\em Journal of Mathematics and Mechanics}, 17(1):59--87, 1967.
\newblock Reprinted in \cite[pp. 235--263]{specker-ges}.

\bibitem{Specker-priv}
Ernst Specker.
\newblock private communication, 1999.

\bibitem{specker-ges}
Ernst Specker.
\newblock {\em Selecta}.
\newblock Birkh{\"{a}}user Verlag, Basel, 1990.

\end{thebibliography}

\end{document}